\documentclass[showpacs,prl,10pt,twocolumn,superscriptaddress,aps]{revtex4-1}
\usepackage{ae}
\usepackage{bm} 
\usepackage{color}
\usepackage{amsmath}
\usepackage{amssymb}
\usepackage{amsfonts}
\usepackage{graphicx}
\usepackage{slashed}
\usepackage{setspace}
\usepackage{stackrel}
\usepackage{ulem}

\newcommand{\Eqref}[1]{Eq.~\eqref{#1}}

\allowdisplaybreaks

\begin{document}

\setlength{\unitlength}{1mm}
\title{An all-loop result for the strong magnetic field limit of the Heisenberg-Euler effective Lagrangian}
\author{Felix Karbstein}\email{felix.karbstein@uni-jena.de}
\affiliation{Helmholtz-Institut Jena, Fr\"obelstieg 3, 07743 Jena, Germany}
\affiliation{Theoretisch-Physikalisches Institut, Abbe Center of Photonics, \\ Friedrich-Schiller-Universit\"at Jena, Max-Wien-Platz 1, 07743 Jena, Germany}

\date{\today}

\begin{abstract}
 We provide an explicit expression for the strong magnetic field limit of the Heisenberg-Euler effective Lagrangian for both scalar and spinor quantum electrodynamics.
 To this end, we show that the strong magnetic field behavior is fully determined by one-particle reducible contributions discovered only recently.
 The latter can efficiently be constructed in an essentially algebraic procedure from lower-order one-particle reducible diagrams.
 Remarkably, the leading strong magnetic field behavior of the all-loop Heisenberg-Euler effective Lagrangian only requires input from the one-loop Lagrangian.
 Our result revises previous findings based exclusively on one-particle irreducible contributions.
 In addition we briefly discuss the strong electric field limit and comment on external field QED in the large $N$ limit.
\end{abstract}

\maketitle

\paragraph{Introduction}

The Heisenberg-Euler effective Lagrangian \cite{Heisenberg:1935qt,Weisskopf,Schwinger:1951nm} is a central quantity in the development of quantum field theory.
It studies the effect of quantum fluctuations on the effective theory of prescribed electromagnetic fields in the vacuum, and allows for the systematic derivation of quantum corrections to Maxwell's classical theory of electrodynamics.
The latter manifest themselves in effective, nonlinear self-couplings between electromagnetic fields, giving rise to light-by-light scattering phenomena \cite{Euler:1935zz}.
Moreover, in electric fields the Heisenberg-Euler effective Lagrangian develops a non-perturbative imaginary part which can be associated with an instability of the quantum vacuum towards the formation of a state featuring real electrons and positrons \cite{Sauter:1931zz}.
For reviews emphasizing various theoretical aspects as well as prospects for the experimental detection of such effects, see \cite{Dittrich:1985yb,Fradkin:1991zq,Dittrich:2000zu,Dunne:2004nc,Marklund:2008gj,Dunne:2008kc,Heinzl:2008an,DiPiazza:2011tq,Dunne:2012vv,Battesti:2012hf,King:2015tba,Karbstein:2016hlj,Battesti:2018bgc}.

Here, we focus on the on-shell renormalized Heisenberg-Euler effective Lagrangian ${\cal L}_{\rm HE}$ for quantum electrodynamics (QED) in $d=3+1$ space-time dimensions and a prescribed constant electromagnetic field $F^{\mu\nu}$.
As in the derivation of ${\cal L}_{\rm HE}$ the dynamical fermion and photon fields are integrated out, ${\cal L}_{\rm HE}$ can be represented in terms of Feynman diagrams featuring internal fermion and photon lines only; cf., e.g., \cite{Gies:2016yaa}.
In turn, the only physical dimensionful scale inherited by ${\cal L}_{\rm HE}$ from the microscopic theory of QED is the electron mass $m$.
Each coupling of a photon to a fermion line is mediated by the elementary charge $e=\sqrt{4\pi\alpha}$, implying that each internal photon line comes with a factor of $\alpha$.
On the other hand, the external field dependence of any loop diagram is entirely in terms of the combined parameter $eF^{\mu\nu}$. The latter combination actually forms a renormalization group (RG) invariant, and hence is independent of the renormalization scale \cite{Ritus:1975}. 
Correspondingly, it is convenient to formally treat the parameter $\alpha$ and the combination $eF^{\mu\nu}$ as independent. The power of the former counts the number of internal photon lines $n_\gamma$ in a given diagram, and the power of the latter the number of couplings to the external fields.

The entire field dependence of ${\cal L}_{\rm HE}$ in constant fields can be encoded in the gauge and Lorentz invariants of the electromagnetic field, ${\cal F}=\frac{1}{4}F_{\mu\nu}F^{\mu\nu}=\frac{1}{2}(\vec{B}^2-\vec{E}^2)$ and ${\cal G}=\frac{1}{4}F_{\mu\nu}{}^\star\!F^{\mu\nu}=-\vec{B}\cdot\vec{E}$, with dual field strength tensor ${}^\star\!F^{\mu\nu}$. Besides, CP invariance of QED demands the effective Lagrangian to be even in the pseudoscalar quantity $\cal G$.
For field configurations fulfilling ${\cal G}=0$ and ${\cal F}>0$ (${\cal F}<0$) it is always possible to find a Lorentz frame in which the field is purely magnetic (electric).
Moreover, a result determined in purely magnetic fields $B=|\vec{B}|$ can always be translated to the analogous one in purely electric fields $E=|\vec{E}|$ by means of the replacement $B\to-{\rm i}E$ and vice versa; cf., e.g., \cite{Jentschura:2001qr}.

The Heisenberg-Euler effective Lagrangian admits a diagrammatic expansion in the number of loops $\ell$ of the constituting Feynman diagrams,
\begin{equation}
 {\cal L}_\text{HE}=\sum_{\ell=0}^\infty {\cal L}^{\ell\text{-loop}}, \label{eq:LHE}
\end{equation}
where ${\cal L}^{0\text{-loop}}=-\frac{1}{4}F_{\mu\nu}F^{\mu\nu}$ is the classical Maxwell Lagrangian.
Note that in purely magnetic fields we have ${\cal L}^{0\text{-loop}}(B)=-\frac{1}{2}B^2$.
The $\ell$-loop contribution scales as ${\cal L}^{\ell\text{-loop}}\sim(\frac{\alpha}{\pi})^{\ell-1}$, where $\alpha\equiv\alpha(m^2)=\frac{e^2}{4\pi}\simeq\frac{1}{137}$ is the fine-structure constant; we use the Heaviside-Lorentz System with $c=\hbar=1$. 
Contributions beyond one loop generically decompose into one-particle irreducible (1PI) and one-particle reducible (1PR) diagrams \cite{Gies:2016yaa}, such that for $\ell\geq2$ we have ${\cal L}^{\ell\text{-loop}}={\cal L}_{1{\rm PI}}^{\ell\text{-loop}}+{\cal L}_{1{\rm PR}}^{\ell\text{-loop}}$; of course, the one-loop Lagrangian ${\cal L}^{1\text{-loop}}$ is 1PI by definition.

The one- and two-loop results ${\cal L}^{1\text{-loop}}$ and ${\cal L}^{2\text{-loop}}$ are known explicitly for both spinor \cite{Heisenberg:1935qt,Ritus:1975,Gies:2016yaa} and scalar \cite{Weisskopf,Ritus:1977} QED; see also \cite{Dittrich:1985yb,Fliegner:1997ra,Kors:1998ew,Dunne:2004nc,Edwards:2017bte,Ahmadiniaz:2017rrk}. 
On the three-loop level, first analytical results for ${\cal L}^{3\text{-loop}}$ have been obtained in 1+1 dimensions \cite{Huet:2009cy,Huet:2011kd,Huet:2018ksz}.

Following the discovery of the nonvanishing of the 1PR contributions to ${\cal L}_{\rm HE}$ in constant fields, a variety of further studies focusing on 1PR contributions in external-field QED have been performed \cite{Edwards:2017bte,Karbstein:2017gsb,Ahmadiniaz:2017rrk,Edwards:2018vjd,Ahmadiniaz:2019nhk}.

\paragraph{Strong magnetic fields}

A particularly interesting parameter regime is the regime of strong magnetic fields characterized by $\frac{eB}{m^2}\gg1$.
In this limit analytical insights are possible at all loop orders.
It is in particular well known in the literature that the 1PI contribution to ${\cal L}_{\rm HE}$ at $\ell$ loops scales as \cite{Weisskopf,Ritus:1975,Dittrich:1975au,Dittrich:1985yb,Heyl:1996dt,Ritus:1998jm,Dittrich:2000zu,Dunne:2004nc}
\begin{align}
  {\cal L}^{1\text{-loop}}(B)=\,&(eB)^2 \frac{\beta_1}{8\pi} \ln\Bigl(\frac{eB}{m^2}\Bigr)\Bigl[1+{\cal O}\bigl(\tfrac{1}{\ln(\frac{eB}{m^2})}\bigr)\Bigr] \quad\text{and} \nonumber\\
  {\cal L}_{1\text{PI}}^{\ell\text{-loop}}(B)=\,& (eB)^2 \frac{\beta_1}{8\pi}\,\frac{\beta_2/\beta_1^{2}}{\ell-1} \Bigl[\alpha\beta_1\ln\Bigl(\frac{eB}{m^2}\Bigr)\Bigr]^{\ell-1} \nonumber\\
  &\quad\times\Bigl[1+{\cal O}\bigl(\tfrac{1}{\ln(\frac{eB}{m^2})}\bigr)\Bigr] \quad\text{for}\quad\ell\geq2\,, \label{eq:1PIsf}
\end{align}
where $\beta_1=\frac{1}{3\pi}$ ($\beta_1=\frac{1}{12\pi}$) and $\beta_2=\frac{1}{4\pi^2}$ ($\beta_2=\frac{1}{4\pi^2}$) 
are the renormalization scheme independent coefficients of the $\beta$ function of spinor (scalar) QED,
\begin{equation}
 \beta\bigl(\alpha(\mu^2)\bigr)=\frac{1}{\alpha(\mu^2)}\,\mu^2\frac{{\rm d}\alpha(\mu^2)}{{\rm d}\mu^2}\,,
\end{equation}
with $\beta\bigl(\alpha\bigr)=\beta_1\alpha+\beta_2\alpha^2+{\cal O}(\alpha^3)$,
governing the running of the fine structure constant.

The structure of \Eqref{eq:1PIsf} can be derived \cite{Ritus:1975} with the help of the Callan-Symanzik equation \cite{Callan:1970yg,Symanzik:1970rt}.
It is the result of a close connection between the short-distance behavior of renormalized Green's functions and the strong-field limit of associated quantities calculated in prescribed background fields \cite{Coleman:1973jx,Ritus:1975,Ritus:1977,Matinyan:1976mp,Dittrich:1985yb,Ritus:1998jm,Dittrich:2000zu,Dunne:2004nc} in the absence of zero modes \cite{Dunne:2002ta}; cf. in particular the review \cite{Dunne:2004nc} and references therein. 

On the other hand, Ref.~\cite{Karbstein:2017gsb} provides an explicit prescription of how to construct all possible 1PR contributions to a given quantity in constant electromagnetic fields from 1PI contributions of lower loop order. 
Adopting the prescription of \cite{Karbstein:2017gsb} to ${\cal L}_{\rm HE}$ in purely magnetic fields, up to potential symmetry factors associated with different diagram topologies which turn out to be irrelevant for the following discussion, we obtain
\begin{align}
 {\cal L}_{1{\rm PR}}^{\ell\text{-loop}}(B)=\,&\frac{1}{2}\sum_{k=1}^{\ell-1} \,
 \sum_{l_1m_1+\cdots+l_nm_n=\ell-k} 
 \Bigl(\frac{1}{2}\frac{{\cal L}_{1\text{PI}}^{l_1\text{-loop}}}{\partial B}\frac{\partial}{\partial B}\Bigr)^{m_1} \nonumber\\
 &\quad\cdots\Bigl(\frac{1}{2}\frac{{\cal L}_{1\text{PI}}^{l_n\text{-loop}}}{\partial B}\frac{\partial}{\partial B}\Bigr)^{m_n} {\cal L}_{1{\rm PI}}^{k\text{-loop}}(B) \,, \label{eq:gellloop}
\end{align}
where the second sum is taken over all sequences of positive integer indices $\{l_i,m_i,n\}\geq1$, with $i\in\{1,\ldots,n\}$, such that $\sum_{i=1}^n l_im_i=\ell-k$.

\begin{figure}
\center
\includegraphics[width=0.29\textwidth]{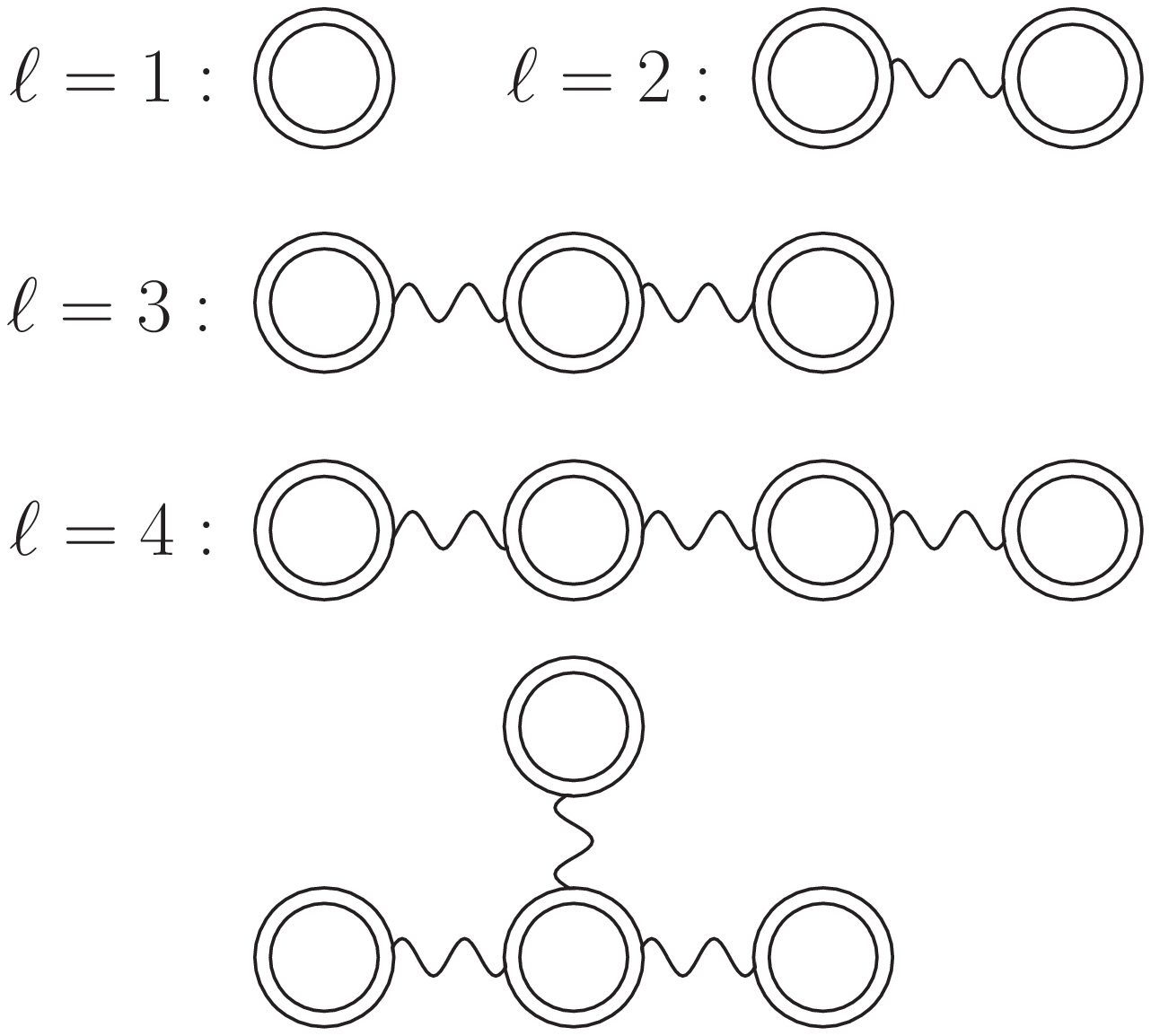}
\caption{Topologically distinct diagrams constituting $\Delta{\cal L}^{\ell\text{-loop}}$ up to four loops.
By definition, all contributing diagrams beyond one loop are one-particle reducible.
The double solid line denotes the charged-particle propagator dressed to all orders by the background field.}
\label{fig:1loopPRdiags}
\end{figure}

As will be shown in a moment, the leading contribution to \Eqref{eq:gellloop} in the strong magnetic field limit arises from contributions involving derivatives of ${\cal L}^{1\text{-loop}}$ only.
Correspondingly, the constituting diagrams are made up of $\ell_f=\ell$ charged-particle loops, and can be expressed as
\begin{equation}
 \Delta{\cal L}^{\ell\text{-loop}}_{1{\rm PR}}(B)=\frac{1}{2}\Bigl(\frac{1}{2}\frac{\partial{\cal L}^{1\text{-loop}}}{\partial B}\frac{\partial}{\partial B}\Bigr)^{\ell-1}{\cal L}^{1\text{-loop}}(B)\,. \label{eq:deltaLB}
\end{equation} 
For a graphical representation, see Fig.~\ref{fig:1loopPRdiags}.

\begin{figure}
\center
\includegraphics[width=0.35\textwidth]{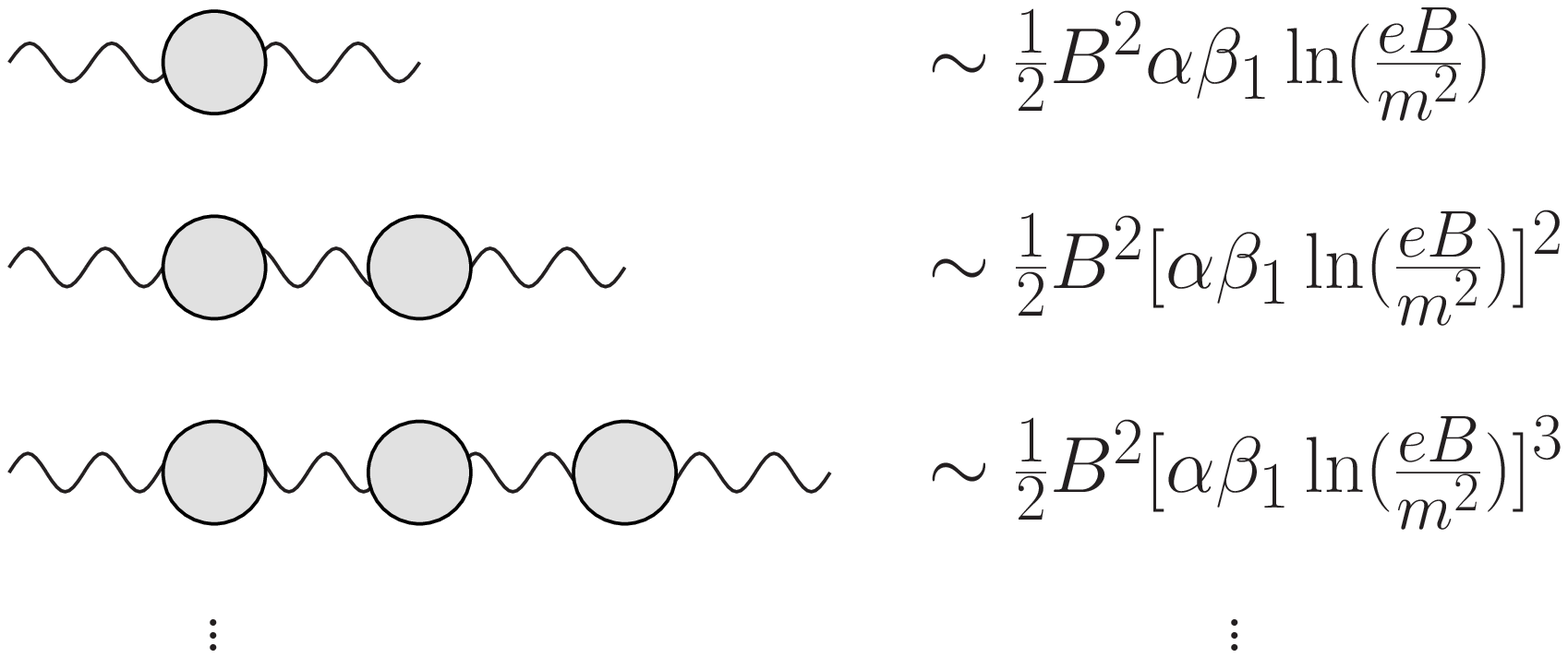}
\caption{In the strong field limit the one-loop Lagrangian~\eqref{eq:1PIsf} effectively acts as quadratic in $B$ (open wiggly lines).
The logarithm can be considered as an inert factor associated with each charged-particle loop (silver pearl). Each coupling comes with a factor of $\sqrt{\alpha}$.
The leading strong field behavior of $\Delta{\cal L}^{\ell\text{-loop}}(B)$ is encoded in the chain featuring $\ell$ pearls.}
\label{fig:resum}
\end{figure}

In a next step, we insert the first line of \Eqref{eq:1PIsf} into \Eqref{eq:deltaLB}.
This will allow us to determine the leading contribution to $\Delta{\cal L}^{\ell\text{-loop}}(B)$ in the limit of a strong magnetic field.
As $\frac{\partial}{\partial B}[B^n\ln^l(B)]%=nB^{n-1}\ln^l(B)[1+\frac{l}{n}\ln^{-1}(B)]
=(\frac{\partial}{\partial B}B^n)\ln^l(B)[1+{\cal O}(\ln^{-1}(B))]$, with $\{n,l\}\in\mathbb{N}$, the leading strong field behavior arises from the contribution with all derivatives for $B$ acting on the power of $B$ and none on the logarithm.
Hence, with regard to the extraction of the leading strong field behavior of $\Delta{\cal L}^{\ell\text{-loop}}$ via derivatives for $B$, \Eqref{eq:1PIsf} effectively appears as quadratic in $B$. The logarithm rather plays the role of an inert factor associated with each charged-particle loop.
For a graphical illustration, see Fig.~\ref{fig:resum}.
In turn, we obtain
\begin{align}
 \Delta{\cal L}^{\ell\text{-loop}}_{1{\rm PR}}(B)=\,&\frac{\beta_1}{8\pi}\bigl[\ln\bigl(\tfrac{eB}{m^2}\bigr)+{\cal O}(1)\bigr]^\ell \!\Bigl(\alpha\beta_1\frac{B}{2}\!\frac{\partial}{\partial B}\Bigr)^{\ell-1}\frac{(eB)^2}{2} \nonumber\\
 =\,&\frac{1}{2}(eB)^2\frac{\beta_1}{8\pi} \ln\bigl(\tfrac{eB}{m^2}\bigr)\bigl[\alpha\beta_1 \ln\bigl(\tfrac{eB}{m^2}\bigr)\bigr]^{\ell-1} \nonumber\\
 &\quad\times\Bigl[1+{\cal O}\bigl(\tfrac{1}{\ln(\frac{eB}{m^2})}\bigr)\Bigr], \label{eq:1PRsf}
\end{align}
for $\ell\geq2$.
This also explains why the dominant 1PR contribution arises from the 1PR diagram involving ${\cal L}^{1\text{-loop}}$ insertions only:
Insertions of higher-loop 1PI contributions~\eqref{eq:1PIsf} into \Eqref{eq:gellloop}  immediately result in a subleading scaling with respect to the power of the logarithm at a given loop order $\ell$.
Correspondingly, for $\ell\geq2$ the leading strong field behavior of the 1PR contribution ${\cal L}^{\ell\text{-loop}}_{1{\rm PR}}$ is given by the expression in \Eqref{eq:1PRsf}.
Subleading 1PR contributions can, in principle, be determined along the same lines.
However, in general their actual determination is much more complicated.

A comparison of Eqs.~\eqref{eq:1PIsf} and \eqref{eq:1PRsf} unveils that for all loop orders $\ell\geq2$ the 1PR contribution dominates the 1PI one, and the leading strong field limit is dictated by the coefficient $\beta_1$ of the QED $\beta$ function alone.
In turn, for $\ell\geq2$ the leading strong field behavior of the Heisenberg Euler effective Lagrangian is given by 
\begin{align}
 {\cal L}^{\ell\text{-loop}}(B)=\,&\frac{1}{2}(eB)^2\frac{\beta_1}{8\pi}\ln\bigl(\tfrac{eB}{m^2}\bigr)\bigl[\alpha\beta_1 \ln\bigl(\tfrac{eB}{m^2}\Bigr)\bigr]^{\ell-1} \nonumber\\
 &\quad\times\Bigl[1+{\cal O}\bigl(\tfrac{1}{\ln(\frac{eB}{m^2})}\bigr)\Bigr]\,. \label{eq:HEsf}
\end{align}
Note that this behavior is in accordance with Weisskopf's investigation \cite{Weisskopf:1939zz} showing that the logarithmic divergence at $\ell\geq2$ loop order scales at most as the $\ell$th power of the logarithm.
A resummation of \Eqref{eq:HEsf} to all loops via \Eqref{eq:LHE} yields
\begin{align}
 {\cal L}_{\rm HE}(B)=-\frac{1}{2}B^2\,+\,&(eB)^2\frac{\beta_1}{8\pi}\ln\bigl(\tfrac{eB}{m^2}\bigr) \nonumber\\
 &\times\Bigl[1+\alpha^{1\text{-loop}}(eB)\frac{\beta_1}{2}\ln\bigl(\tfrac{eB}{m^2}\bigr)\Bigr] \nonumber\\
 &\times\Bigl[1+{\cal O}\bigl(\tfrac{1}{\ln(\frac{eB}{m^2})}\bigr)\Bigr]\,, \label{eq:LHEsf}
\end{align} 
with the one-loop running of the fine structure given by
\begin{equation}
 \alpha^{1\text{-loop}}(\mu^2)=\frac{\alpha}{1-\alpha\beta_1\ln(\frac{\mu^2}{m^2})}\,, \label{eq:alpha1loop}
\end{equation}
and $\alpha\equiv\alpha(m^2)$.
This expression~\eqref{eq:LHEsf} is convergent for $\alpha\beta_1\ln(eB/m^2)<1$.
The appearance of \Eqref{eq:alpha1loop} in this context is not surprising.
The resummation of chain diagrams as depicted in Fig.~\ref{fig:resum} is reminiscent of the calculation of an exact two-point function by summing up all 1PI insertions.
Upon restriction to one-loop insertions this generically gives rise to the one-loop running of quantities. 

Equation~\eqref{eq:LHEsf} supersedes the expression obtained in \cite{Dittrich:1985yb} which is based on the resummation of the leading logarithms in the 1PI sector of ${\cal L}_{\rm HE}$ only and completely neglects 1PR contributions.
It is interesting to note the structural similarity of this result with the leading contributions to $\sum_{\ell=0}^2{\cal L}^{\ell\text{-loop}}(B)$ given in Eqs.~\eqref{eq:1PIsf} and \eqref{eq:HEsf};
more specifically, we have ${\cal L}_{\rm HE}(B)=-\frac{1}{2}B^2+{\cal L}^{1\text{-loop}}(B)+{\cal L}^{2\text{-loop}}(B)|_{\alpha\to\alpha^{1\text{-loop}}(eB)}$.
This structure is actually not too surprising:
From an effective field theory point of view it is natural that the couplings are evaluated at the relevant momentum scale, which in the strong magnetic field limit $eB\gg m^2$ amounts to the RG invariant combination $eB$.

Some additional comments are in order here. The coupling \eqref{eq:alpha1loop} diverges at the Landau pole $\mu^2=m^2\,{\rm e}^{\frac{1}{\alpha\beta_1}}$.
In the present case, the pole in the all-order result \eqref{eq:LHEsf} cannot be considered as the artifact of a finite-loop-order calculation which might presumably be removed by higher-loop corrections within perturbative QED.
Manifestly non-perturbative lattice studies of QED indicate that it could nevertheless lie in a region of the parameter space which is made inaccessible by spontaneous chiral symmetry breaking \cite{Gockeler:1997dn}.
In any case, within the Standard Model of particle physics quantum corrections of other particle degrees of freedom become relevant way before the Landau pole of QED is reached.

Even though it is clearly beyond the validity regime of perturbation theory, it is interesting to observe that in the formal limit of $\alpha\beta_1\ln(eB/m^2)\gg1$, i.e., beyond the Landau pole, the factor of unity in the denominator of \Eqref{eq:alpha1loop} can be 
neglected and the terms in the second line of \Eqref{eq:LHEsf} combine to a factor of $\frac{1}{2}$ such that ${\cal L}_\text{HE}(B)\to-\frac{1}{2}B^2+\frac{1}{2}(eB)^2\frac{\beta_1}{8\pi}\ln(\frac{eB}{m^2})$.

\paragraph{Strong electric fields}

As noted in the introduction, the result for a purely electric field follows from the purely magnetic field one by means of the replacement $B\to-{\rm i}E$.
Taking into account that $\ln(eB/m^2)|_{B\to-{\rm i}E}=\ln(eE/m^2)-{\rm i}\pi/2$, it is straightforward to infer that
\begin{align}
 \alpha^{1\text{-loop}}(eB)\big|_{B\to-{\rm i}E}&=\frac{\alpha}{1-\alpha\beta_1\ln(\frac{eE}{m^2})} \nonumber\\
 &\quad\quad\times\Bigl[1+{\cal O}\bigl(\tfrac{1}{\ln(\frac{eE}{m^2})}\bigr)\Bigr]\,. \label{eq:alpha1loopE}
\end{align}
In turn, the leading strong electric field behavior of ${\cal L}_\text{HE}$ is given by
\begin{align}
 {\cal L}_{\rm HE}(E)=\frac{1}{2}E^2\,-\,&(eE)^2\frac{\beta_1}{8\pi}\Bigl[\ln\bigl(\tfrac{eE}{m^2}\bigr)-{\rm i}\frac{\pi}{2}\Bigr] \nonumber\\
 &\times\Bigl\{1+\alpha^{1\text{-loop}}(eE)\frac{\beta_1}{2}\Bigl[\ln\bigl(\tfrac{eE}{m^2}\bigr)-{\rm i}\frac{\pi}{2}\Bigr]\Bigr\} \nonumber\\
 &\times\Bigl[1+{\cal O}\bigl(\tfrac{1}{\ln(\frac{eE}{m^2})}\bigr)\Bigr]\,. \label{eq:LHEsfE}
\end{align} 
Note again the structural similarity of the all-order result~\eqref{eq:LHEsfE} and the analogous expression for $\sum_{\ell=1}^2{\cal L}^{\ell\text{-loop}}(E)$. In the considered limit we have ${\cal L}_{\rm HE}(E)=\frac{1}{2}E^2+{\cal L}^{1\text{-loop}}+{\cal L}^{2\text{-loop}}(E)|_{\alpha\to\alpha^{1\text{-loop}}(eE)}$.

From \Eqref{eq:LHEsfE} we can straightforwardly infer the all-loop strong electric field limit of the vacuum decay rate $w(E)=2\,{\rm Im}\{{\cal L}_\text{HE}(E)\}$ \cite{Heisenberg:1935qt,Schwinger:1951nm}; c.f. also \cite{Dittrich:1985yb,Dunne:2004nc,Cohen:2008wz} and references therein.
Its explicit expression is given by
\begin{align}
 w(E)=\,&(eE)^2\frac{\beta_1}{8}\bigl[1+\alpha^{1\text{-loop}}(eE)\,\beta_1\ln\bigl(\tfrac{eE}{m^2}\bigr)\bigr] \nonumber\\
 &\quad\times\Bigl[1+{\cal O}\bigl(\tfrac{1}{\ln(\frac{eE}{m^2})}\bigr)\Bigr]\,. \label{eq:w(E)}
\end{align}
Previously it was believed that the corresponding one-loop result, $w^{1\text{-loop}}(E)=(eE)^2\frac{\beta_1}{8}\bigl[1+{\cal O}\bigl(1/\ln(\frac{eE}{m^2})\bigr)\bigr]$, does not receive corrections from higher loops as long as $\alpha\beta_1\ln(\frac{eE}{m^2})\lesssim1$ \cite{Dittrich:1985yb}.
As $\frac{w(E)}{w^{1\text{-loop}}(E)}\simeq 1+\alpha^{1\text{-loop}}(eE)\,\beta_1\ln(\frac{eE}{m^2})$, these quantities may, however, differ significantly from each other, and the new all-loop result~\eqref{eq:w(E)} in general predicts a larger vacuum decay rate.

\paragraph{External field QED in the large $N$ limit}

In the following, we briefly comment on the structure of the Heisenberg-Euler effective Lagrangian for QED with $N$ charged particle flavors of equal mass $m$. In contrast to standard QED, this theory features $N$ generations of electrons and positrons. More specifically we consider the 't Hooft limit, characterized by sending 
\begin{equation}
 N\to\infty\,,\quad\text{while keeping}\quad N\alpha=\text{const.} \label{eq:'tHooft}
\end{equation}
At first sight, \Eqref{eq:'tHooft} necessarily implies $eF^{\mu\nu}\sim1/\sqrt{N}$, which would render this limit rather uninteresting as physics would then be dominated by processes at zero field.
However, we can easily arrive at an more interesting limit by demanding in addition that
\begin{equation}
 F^{\mu\nu}\sim \sqrt{N}\,,\quad\text{such that}\quad eF^{\mu\nu}=\text{const.} \label{eq:eF=const.}
\end{equation}
As the intensity of the prescribed field scales quadratically with $F^{\mu\nu}$, and the intensity generically scales linear with the number ${\cal N}$ of (microscopically unresolved) photons forming the macroscopic external field, physically \Eqref{eq:eF=const.}
can be associated with the case where ${\cal N}\sim N$.

The effective Lagrangian ${\cal L}_\text{HE}$ for QED with $N$ charged particle flavors of equal mass follows straightforwardly from the one of ordinary QED with $N=1$:
Aiming at extracting the former from the latter, it is helpful to represent it in terms of Feynman diagrams.
As each fermion loop comes with a factor of $N$, the contribution of a Feynman diagram containing $\ell_f$ fermion loops is to be multiplied with an overall factor of $N^{\ell_f}$.
On the other hand, in the 't Hooft limit~\eqref{eq:'tHooft} each internal photon line comes with a factor of $\alpha\sim1/N$.
Hence, a Feynman diagram featuring $\ell_f$ fermion loops and $n_\gamma$ internal photon lines scales as $N^{\ell_f-n_\gamma}$, such that the diagrams maximizing the difference $\ell_f-n_\gamma>0$ dominate in the large $N$ limit.
At a given loop order $\ell$, at most all loops are fermionic ones, i.e., $\ell=\ell_f$.
To connect these $\ell_f$ fermion loops without forming an additional loop $n_\gamma=\ell_f-1$ photon lines are needed, and the respective class of diagrams scales linear with $N$, independently of the loop order.
Given that such diagrams exist, this immediately implies that the leading diagrams at any loop order exhibit the same scaling with $N$ and thus are equally important.

In purely magnetic fields, the latter class of diagrams is precisely generated by \Eqref{eq:deltaLB}, which also governs the leading strong magnetic (electric) field behavior of ordinary QED; cf. also Fig.~\ref{fig:1loopPRdiags}.
For the analogous result in arbitrary constant electromagnetic fields, see Ref.~\cite{Karbstein:2017gsb}.
Correspondingly, the strong magnetic field limit of the Heisenberg-Euler effective Lagrangian in the 't Hooft limit~\eqref{eq:'tHooft} and \eqref{eq:eF=const.} is given by
\begin{align}
 {\cal L}_{\rm HE}^{\text{large}\,N}\!(B)=-\frac{1}{2}B^2\,+\,&(eB)^2\frac{N\beta_1}{8\pi}\ln(\tfrac{eB}{m^2}) \nonumber\\
 &\times\Bigl[1+\frac{1}{2}\frac{\alpha N \beta_1\ln(\tfrac{eB}{m^2})}{1-\alpha N\beta_1\ln(\frac{eB}{m^2})}\Bigr] \nonumber\\
 &\times\Bigl[1+{\cal O}(\tfrac{1}{\ln(\frac{eB}{m^2})})\Bigr]\,. \label{eq:LargeNB}
\end{align} 
The analogous expression for a purely electric field is
\begin{align}
 {\cal L}_{\rm HE}^{\text{large}\,N}\!(E)=\frac{1}{2}E^2\,-\,&(eE)^2\frac{N\beta_1}{8\pi}\Bigl[\ln(\tfrac{eE}{m^2})-{\rm i}\frac{\pi}{2}\Bigr] \nonumber\\
 &\times\Bigl[1+\frac{1}{2}\frac{\alpha N \beta_1\bigl[\ln(\tfrac{eE}{m^2})-{\rm i}\frac{\pi}{2}\bigr]}{1-\alpha N\beta_1\ln(\frac{eE}{m^2})}\Bigr] \nonumber\\
 &\times\Bigl[1+{\cal O}(\tfrac{1}{\ln(\frac{eE}{m^2})})\Bigr]\,. \label{eq:LargeNE}
\end{align}

We emphasize that even though for $N=1$ Eqs.~\eqref{eq:LargeNB} and \eqref{eq:LargeNE} reproduce the ordinary QED results discussed in Eqs.~\eqref{eq:LHEsf} and \eqref{eq:LHEsfE}, they would even be correct when the correspondingly 1PI contributions in \Eqref{eq:1PIsf} would exhibit the same logarithmic scaling as in \Eqref{eq:1PRsf}, the reason being that the 1PI contributions with $\ell\geq2$ loops are generically suppressed by factors of $1/N$. 

Finally, we note that prior to the discovery of the nonvanishing of the 1PR contributions to ${\cal L}_{\rm HE}$ by Ref.~\cite{Gies:2016yaa}, the one-loop result ${\cal L}^{1\text{-loop}}$ had been the only loop diagram which would have been considered as yielding a contribution $\sim N$.
Instead, infinitely many 1PR diagrams contribute at linear order in $N$.

\paragraph{Conclusions and Outlook}

In this article, we have explicitly determined the strong magnetic field limit of the Heisenberg-Euler effective Lagrangian.
After demonstrating that beyond one loop this limit is fully determined by one-particle reducible contributions which were discovered only recently \cite{Gies:2016yaa}, we extracted the leading contribution at each loop order $\ell$.
In a next step, we resummed these leading contributions to obtain the leading strong magnetic field behavior of the all-loop Heisenberg-Euler effective Lagrangian.
This result could then be straightforwardly translated to the case of a purely electric field, by means of the replacement $B\to-{\rm i}E$.

Finally, we briefly commented on external field QED in the large $N$ 't Hooft limit.
Here, we  emphasized in particular the fact that the leading large $N$ behavior of the Heisenberg-Euler effective Lagrangian receives contributions from infinitely many one-particle reducible diagrams.
The latter can, however, be efficiently constructed in an essentially algebraic procedure from the one-loop Heisenberg-Euler Lagrangian.
This facilitates unprecedented analytical studies of the large $N$ all-loop Heisenberg-Euler Lagrangian at arbitrary field strengths.  

Focusing on the paradigmatic example of the strong electric (magnetic) field limit of the Heisenberg-Euler effective Lagrangian, our results clearly illustrate that the whole class of 1PR tadpole diagrams in constant external fields \cite{Karbstein:2017gsb}, which until recently \cite{Gies:2016yaa} where believed to vanish identically, are not only non-zero but may even dominate physical observables of interest.

We expect our findings to be of interest not only for the research field of external-field QED, but also for other field theories in prescribed external fields, such as quantum chromodynamics (QCD) in (gluo)magnetic background fields where constant-field 1PR tadpole contributions were so-far not accounted for.

\acknowledgments

It is a great pleasure to thank Holger Gies for valuable comments on this manuscript.
This work has been funded by the Deutsche Forschungsgemeinschaft (DFG) under Grant No. 416607684 within the Research Unit FOR2783/1.


\begin{thebibliography}{10}\setlength{\itemsep}{-0.5mm}

%\cite{Heisenberg:1935qt}
\bibitem{Heisenberg:1935qt} 
  W.~Heisenberg and H.~Euler,
  %``Folgerungen aus der Diracschen Theorie des Positrons,''
  Z.\ Phys.\  {\bf 98}, 714 (1936), 
  an English translation is available at [physics/0605038].
  %%CITATION = PHYSICS/0605038;%%
  %%CITATION = ZEPYA,98,714;%%
  %879 citations counted in INSPIRE as of 06 May 2013

%\cite{Weisskopf}
\bibitem{Weisskopf}
  V.~Weisskopf,
  %``\"Uber die Elektrodynamik des Vakuums auf Grund der Quanthentheorie des Elektrons,''
  Kong.\ Dans.\ Vid.\ Selsk., Mat.-fys.\ Medd.\ {\bf XIV}, 6 (1936).

%\cite{Schwinger:1951nm}
\bibitem{Schwinger:1951nm} 
  J.~S.~Schwinger,
  %``On gauge invariance and vacuum polarization,''
  Phys.\ Rev.\ {\bf 82}, 664 (1951).
  %%CITATION = PHRVA,82,664;%%
  %3762 citations counted in INSPIRE as of 09 Oct 2015

%\cite{Euler:1935zz}
\bibitem{Euler:1935zz} 
  H.~Euler and B.~Kockel,
  %``\"Uber die Streuung von Licht an Licht nach der Diracschen Theorie,''
  Naturwiss.\  {\bf 23}, 246 (1935).
  %%CITATION = NATWA,23,246;%%
  %137 citations counted in INSPIRE as of 23 Feb 2015
 

%\cite{Sauter:1931zz}
\bibitem{Sauter:1931zz} 
  F.~Sauter,
  %``Uber das Verhalten eines Elektrons im homogenen elektrischen Feld nach der relativistischen Theorie Diracs,''
  Z.\ Phys.\  {\bf 69}, 742 (1931).
  %doi:10.1007/BF01339461
  %%CITATION = doi:10.1007/BF01339461;%%
  %497 citations counted in INSPIRE as of 04 Mar 2019
 
%\cite{Dittrich:1985yb}
\bibitem{Dittrich:1985yb} 
  W.~Dittrich and M.~Reuter,
  %``Effective Lagrangians In Quantum Electrodynamics,''
  Lect.\ Notes Phys.\  {\bf 220}, 1 (1985).
  %%CITATION = LNPHA,220,1;%%
  %91 citations counted in INSPIRE as of 22 Feb 2019

%\cite{Fradkin:1991zq}
\bibitem{Fradkin:1991zq} 
  E.~S.~Fradkin, D.~M.~Gitman and S.~M.~Shvartsman,
  %``Quantum electrodynamics with unstable vacuum,''
  Berlin, Germany: Springer (1991) 288 p. (Springer series in nuclear and particle physics)
  %37 citations counted in INSPIRE as of 04 Mar 2019
  
%\cite{Dittrich:2000zu}
\bibitem{Dittrich:2000zu} 
  W.~Dittrich and H.~Gies,
  %``Probing the quantum vacuum. Perturbative effective action approach in quantum electrodynamics and its application,''
  Springer Tracts Mod.\ Phys.\  {\bf 166}, 1 (2000).
  %%CITATION = STPHB,166,1;%%
  %99 citations counted in INSPIRE as of 06 May 2013
 
%\cite{Dunne:2004nc}
\bibitem{Dunne:2004nc} 
  G.~V.~Dunne,
  %``Heisenberg-Euler effective Lagrangians: Basics and extensions,''
  In *Shifman, M. (ed.) et al.: From fields to strings, vol. 1* 445-522
  [hep-th/0406216].
  %%CITATION = HEP-TH/0406216;%%
  %187 citations counted in INSPIRE as of 19 août 2015
 
%\cite{Marklund:2008gj}
\bibitem{Marklund:2008gj} 
  M.~Marklund and J.~Lundin,
  %``Quantum Vacuum Experiments Using High Intensity Lasers,''
  Eur.\ Phys.\ J.\ D {\bf 55}, 319 (2009)
  [arXiv:0812.3087 [hep-th]].
  %%CITATION = ARXIV:0812.3087;%%
  %18 citations counted in INSPIRE as of 06 May 2013

%\cite{Dunne:2008kc}
\bibitem{Dunne:2008kc} 
  G.~V.~Dunne,
  %``New Strong-Field QED Effects at ELI: Nonperturbative Vacuum Pair Production,''
  Eur.\ Phys.\ J.\ D {\bf 55}, 327 (2009)
  [arXiv:0812.3163 [hep-th]].
  %%CITATION = ARXIV:0812.3163;%%
  %45 citations counted in INSPIRE as of 06 May 2013

%\cite{Heinzl:2008an}
\bibitem{Heinzl:2008an} 
  T.~Heinzl and A.~Ilderton,
  %``Exploring high-intensity QED at ELI,''
  Eur.\ Phys.\ J.\ D {\bf 55}, 359 (2009)
  [arXiv:0811.1960 [hep-ph]].
  
%\cite{DiPiazza:2011tq}
\bibitem{DiPiazza:2011tq} 
  A.~Di Piazza, C.~Muller, K.~Z.~Hatsagortsyan and C.~H.~Keitel,
  %``Extremely high-intensity laser interactions with fundamental quantum systems,''
  Rev.\ Mod.\ Phys.\  {\bf 84}, 1177 (2012)
  [arXiv:1111.3886 [hep-ph]].
  %%CITATION = ARXIV:1111.3886;%%
  %34 citations counted in INSPIRE as of 06 May 2013

%\cite{Dunne:2012vv}
\bibitem{Dunne:2012vv} 
  G.~V.~Dunne,
  %``The Heisenberg-Euler Effective Action: 75 years on,''
  Int.\ J.\ Mod.\ Phys.\ A {\bf 27}, 1260004 (2012)
  [Int.\ J.\ Mod.\ Phys.\ Conf.\ Ser.\  {\bf 14}, 42 (2012)]
  [arXiv:1202.1557 [hep-th]].
  %%CITATION = ARXIV:1202.1557;%%
  %9 citations counted in INSPIRE as of 12 Jan 2015
  
%\cite{Battesti:2012hf}
\bibitem{Battesti:2012hf} 
  R.~Battesti and C.~Rizzo,
  %``Magnetic and electric properties of quantum vacuum,''
  Rept.\ Prog.\ Phys.\  {\bf 76}, 016401 (2013)
  [arXiv:1211.1933 [physics.optics]].
  %%CITATION = ARXIV:1211.1933;%%
  %13 citations counted in INSPIRE as of 01 Dec 2014
  
%\cite{King:2015tba}
\bibitem{King:2015tba} 
  B.~King and T.~Heinzl,
  %``Measuring Vacuum Polarisation with High Power Lasers,''
  High Power Laser Science and Engineering, 4, e5 (2016)
  %doi:10.1017/hpl.2016.1
  [arXiv:1510.08456 [hep-ph]].
  %%CITATION = doi:10.1017/hpl.2016.1;%%
  %8 citations counted in INSPIRE as of 21 Nov 2016
 
%\cite{Karbstein:2016hlj}
\bibitem{Karbstein:2016hlj} 
  F.~Karbstein,
  %``The quantum vacuum in electromagnetic fields: From the Heisenberg-Euler effective action to vacuum birefringence,''
  %doi:10.3204/DESY-PROC-2016-04
  arXiv:1611.09883 [hep-th].
  %%CITATION = doi:10.3204/DESY-PROC-2016-04;%%

%\cite{Battesti:2018bgc}
\bibitem{Battesti:2018bgc} 
  R.~Battesti {\it et al.},
  %``High magnetic fields for fundamental physics,''
  Phys.\ Rept.\  {\bf 765-766}, 1 (2018)
  %doi:10.1016/j.physrep.2018.07.005
  [arXiv:1803.07547 [physics.ins-det]].
  %%CITATION = doi:10.1016/j.physrep.2018.07.005;%%
  %2 citations counted in INSPIRE as of 10 Mar 2019

%\cite{Gies:2016yaa}
\bibitem{Gies:2016yaa} 
  H.~Gies and F.~Karbstein,
  %``An Addendum to the Heisenberg-Euler effective action beyond one loop,''
  JHEP {\bf 1703}, 108 (2017)
  %doi:10.1007/JHEP03(2017)108
  [arXiv:1612.07251 [hep-th]].
  %%CITATION = doi:10.1007/JHEP03(2017)108;%%
  %6 citations counted in INSPIRE as of 18 Jun 2017
  
%\cite{Ritus:1975} 
\bibitem{Ritus:1975} 
  V.~I.~Ritus,
  %``Lagrangian of an intense electromagnetic field and quantum electrodynamics at short distances,''
  Zh.\ Eksp.\ Teor.\ Fiz.\  {\bf 69}, 1517 (1975)\ [Sov.\ Phys.\ JETP\ {\bf 42}, 774 (1975)].
 
%\cite{Jentschura:2001qr}
\bibitem{Jentschura:2001qr} 
  U.~D.~Jentschura, H.~Gies, S.~R.~Valluri, D.~R.~Lamm and E.~J.~Weniger,
  %``QED effective action revisited,''
  Can.\ J.\ Phys.\  {\bf 80}, 267 (2002)
  %doi:10.1139/p01-139
  [hep-th/0107135].
  %%CITATION = doi:10.1139/p01-139;%%
  %17 citations counted in INSPIRE as of 04 Mar 2019
 
\bibitem{Ritus:1977} 
  V.~I.~Ritus,
  %``Connection between strong-field quantum electrodynamics with shortdistance quantum electrodynamics,''
  Zh.\ Eksp.\ Teor.\ Fiz.\  {\bf 73}, 807 (1977)\ [Sov.\ Phys.\ JETP\ {\bf 46}, 423 (1977)].

%\cite{Fliegner:1997ra}
\bibitem{Fliegner:1997ra} 
  D.~Fliegner, M.~Reuter, M.~G.~Schmidt and C.~Schubert,
  %``The Two loop Euler-Heisenberg Lagrangian in dimensional renormalization,''
  Theor.\ Math.\ Phys.\  {\bf 113}, 1442 (1997)
  [Teor.\ Mat.\ Fiz.\  {\bf 113}, 289 (1997)]
  %doi:10.1007/BF02634170
  [hep-th/9704194].
  %%CITATION = doi:10.1007/BF02634170;%%
  %53 citations counted in INSPIRE as of 18 Dec 2016

%\cite{Kors:1998ew}
\bibitem{Kors:1998ew} 
  B.~Kors and M.~G.~Schmidt,
  %``The Effective two loop Euler-Heisenberg action for scalar and spinor QED in a general constant background field,''
  Eur.\ Phys.\ J.\ C {\bf 6}, 175 (1999)
  %doi:10.1007/s100500050332, 10.1007/s100520050331
  [hep-th/9803144].
  %%CITATION = doi:10.1007/s100500050332, 10.1007/s100520050331;%%
  %43 citations counted in INSPIRE as of 18 Dec 2016
 
%\cite{Edwards:2017bte}
\bibitem{Edwards:2017bte} 
  J.~P.~Edwards and C.~Schubert,
  %``One-particle reducible contribution to the one-loop scalar propagator in a constant field,''
  Nucl.\ Phys.\ B {\bf 923}, 339 (2017)
  %doi:10.1016/j.nuclphysb.2017.08.002
  [arXiv:1704.00482 [hep-th]].
  %%CITATION = doi:10.1016/j.nuclphysb.2017.08.002;%%
  %1 citations counted in INSPIRE as of 12 Sep 2017
  
%\cite{Ahmadiniaz:2017rrk}
\bibitem{Ahmadiniaz:2017rrk} 
  N.~Ahmadiniaz, F.~Bastianelli, O.~Corradini, J.~P.~Edwards and C.~Schubert,
  %``One-particle reducible contribution to the one-loop spinor propagator in a constant field,''
  Nucl.\ Phys.\ B {\bf 924}, 377 (2017)
  %doi:10.1016/j.nuclphysb.2017.09.012
  [arXiv:1704.05040 [hep-th]].
  %%CITATION = doi:10.1016/j.nuclphysb.2017.09.012;%%
  %9 citations counted in INSPIRE as of 24 Feb 2019
  
%\cite{Huet:2009cy}
\bibitem{Huet:2009cy} 
  I.~Huet, D.~G.~C.~McKeon and C.~Schubert,
  %``Three-loop Euler-Heisenberg Lagrangian and asymptotic analysis in 1+1 QED,''
  %doi:10.1142/9789814289931\_0064
  arXiv:0911.0227 [hep-th].
  %%CITATION = doi:10.1142/9789814289931_0064;%%
  %2 citations counted in INSPIRE as of 16 Dec 2016

%\cite{Huet:2011kd}
\bibitem{Huet:2011kd} 
  I.~Huet, M.~Rausch de Traubenberg and C.~Schubert,
  %``The Euler-Heisenberg Lagrangian Beyond One Loop,''
  Int.\ J.\ Mod.\ Phys.\ Conf.\ Ser.\  {\bf 14}, 383 (2012)
%  doi:10.1142/S2010194512007507
  [arXiv:1112.1049 [hep-th]].
  %%CITATION = doi:10.1142/S2010194512007507;%%
  %4 citations counted in INSPIRE as of 16 Dec 2016

%\cite{Huet:2018ksz}
\bibitem{Huet:2018ksz} 
  I.~Huet, M.~Rausch De Traubenberg and C.~Schubert,
  %``Three-loop Euler-Heisenberg Lagrangian in 1$+$1 QED, part 1: single fermion-loop part,''
  JHEP {\bf 1903}, 167 (2019)
  %doi:10.1007/JHEP03(2019)167
  [arXiv:1812.08380 [hep-th]].
  %%CITATION = doi:10.1007/JHEP03(2019)167;%%
  %2 citations counted in INSPIRE as of 22 May 2019

%\cite{Karbstein:2017gsb}
\bibitem{Karbstein:2017gsb} 
  F.~Karbstein,
  %``Tadpole diagrams in constant electromagnetic fields,''
  JHEP {\bf 1710}, 075 (2017)
  %doi:10.1007/JHEP10(2017)075
  [arXiv:1709.03819 [hep-th]].
  %%CITATION = doi:10.1007/JHEP10(2017)075;%%
  %5 citations counted in INSPIRE as of 21 Feb 2019
  
%\cite{Edwards:2018vjd}
\bibitem{Edwards:2018vjd} 
  J.~P.~Edwards, A.~Huet and C.~Schubert,
  %``On the low-energy limit of the QED N-photon amplitudes: part 2,''
  Nucl.\ Phys.\ B {\bf 935}, 198 (2018)
  %doi:10.1016/j.nuclphysb.2018.07.026
  [arXiv:1807.10697 [hep-th]].
  %%CITATION = doi:10.1016/j.nuclphysb.2018.07.026;%%
  %2 citations counted in INSPIRE as of 24 Feb 2019
  
%\cite{Ahmadiniaz:2019nhk}
\bibitem{Ahmadiniaz:2019nhk} 
  N.~Ahmadiniaz, J.~P.~Edwards and A.~Ilderton,
  %``Reducible contributions to quantum electrodynamics in external fields,''
  JHEP {\bf 1905}, 038 (2019)
  %doi:10.1007/JHEP05(2019)038
  [arXiv:1901.09416 [hep-th]].
  %%CITATION = doi:10.1007/JHEP05(2019)038;%%
  %1 citations counted in INSPIRE as of 22 May 2019

  
%\cite{Dittrich:1975au}
\bibitem{Dittrich:1975au} 
  W.~Dittrich,
  %``One Loop Effective Potentials in QED,''
  J.\ Phys.\ A {\bf 9}, 1171 (1976).
  %doi:10.1088/0305-4470/9/7/019
  %%CITATION = doi:10.1088/0305-4470/9/7/019;%%
  %22 citations counted in INSPIRE as of 22 Feb 2019
  
%\cite{Heyl:1996dt}
\bibitem{Heyl:1996dt} 
  J.~S.~Heyl and L.~Hernquist,
  %``An Analytic form for the effective Lagrangian of QED and its application to pair production and photon splitting,''
  Phys.\ Rev.\ D {\bf 55}, 2449 (1997)
  %doi:10.1103/PhysRevD.55.2449
  [hep-th/9607124].
  %%CITATION = doi:10.1103/PhysRevD.55.2449;%%
  %58 citations counted in INSPIRE as of 22 Feb 2019  

%\cite{Ritus:1998jm}
\bibitem{Ritus:1998jm} 
  V.~I.~Ritus,
  %``Effective Lagrange function of intense electromagnetic field in QED,''
  In *Sandansky 1998, Frontier tests of QED and physics of the vacuum* 11-28
  [hep-th/9812124].
  %%CITATION = HEP-TH/9812124;%%
  %25 citations counted in INSPIRE as of 24 Feb 2019

%\cite{Callan:1970yg}
\bibitem{Callan:1970yg} 
  C.~G.~Callan, Jr.,
  %``Broken scale invariance in scalar field theory,''
  Phys.\ Rev.\ D {\bf 2}, 1541 (1970).
  %doi:10.1103/PhysRevD.2.1541
  %%CITATION = doi:10.1103/PhysRevD.2.1541;%%
  %1043 citations counted in INSPIRE as of 25 Feb 2019

%\cite{Symanzik:1970rt}
\bibitem{Symanzik:1970rt} 
  K.~Symanzik,
  %``Small distance behavior in field theory and power counting,''
  Commun.\ Math.\ Phys.\  {\bf 18}, 227 (1970).
  %doi:10.1007/BF01649434
  %%CITATION = doi:10.1007/BF01649434;%%
  %1039 citations counted in INSPIRE as of 25 Feb 2019
  
%\cite{Coleman:1973jx}
\bibitem{Coleman:1973jx} 
  S.~R.~Coleman and E.~J.~Weinberg,
  %``Radiative Corrections as the Origin of Spontaneous Symmetry Breaking,''
  Phys.\ Rev.\ D {\bf 7}, 1888 (1973).
  %doi:10.1103/PhysRevD.7.1888
  %%CITATION = doi:10.1103/PhysRevD.7.1888;%%
  %4200 citations counted in INSPIRE as of 24 Feb 2019

%\cite{Matinyan:1976mp}
\bibitem{Matinyan:1976mp} 
  S.~G.~Matinyan and G.~K.~Savvidy,
  %``Vacuum Polarization Induced by the Intense Gauge Field,''
  Nucl.\ Phys.\ B {\bf 134}, 539 (1978).
  %doi:10.1016/0550-3213(78)90463-7
  %%CITATION = doi:10.1016/0550-3213(78)90463-7;%%
  %320 citations counted in INSPIRE as of 24 Feb 2019

%\cite{Dunne:2002ta}
\bibitem{Dunne:2002ta} 
  G.~V.~Dunne, H.~Gies and C.~Schubert,
  %``Zero modes, beta functions and IR / UV interplay in higher loop QED,''
  JHEP {\bf 0211}, 032 (2002)
  %doi:10.1088/1126-6708/2002/11/032
  [hep-th/0210240].
  %%CITATION = doi:10.1088/1126-6708/2002/11/032;%%
  %27 citations counted in INSPIRE as of 16 Mar 2019

%\cite{Weisskopf:1939zz}
\bibitem{Weisskopf:1939zz} 
  V.~F.~Weisskopf,
  %``On the Self-Energy and the Electromagnetic Field of the Electron,''
  Phys.\ Rev.\  {\bf 56}, 72 (1939).
  %doi:10.1103/PhysRev.56.72
  %%CITATION = doi:10.1103/PhysRev.56.72;%%
  %146 citations counted in INSPIRE as of 25 Feb 2019
 
%\cite{Gockeler:1997dn}
\bibitem{Gockeler:1997dn} 
  M.~G\"ockeler, R.~Horsley, V.~Linke, P.~E.~L.~Rakow, G.~Schierholz and H.~Stuben,
  %``Is there a Landau pole problem in QED?,''
  Phys.\ Rev.\ Lett.\  {\bf 80}, 4119 (1998)
  %doi:10.1103/PhysRevLett.80.4119
  [hep-th/9712244].
  %%CITATION = doi:10.1103/PhysRevLett.80.4119;%%
  %64 citations counted in INSPIRE as of 28 Apr 2019

 
%\cite{Cohen:2008wz}
\bibitem{Cohen:2008wz} 
  T.~D.~Cohen and D.~A.~McGady,
  %``The Schwinger mechanism revisited,''
  Phys.\ Rev.\ D {\bf 78}, 036008 (2008)
  %doi:10.1103/PhysRevD.78.036008
  [arXiv:0807.1117 [hep-ph]].
  %%CITATION = doi:10.1103/PhysRevD.78.036008;%%
  %64 citations counted in INSPIRE as of 27 Apr 2019
 
\end{thebibliography}
\end{document}